# The IoT Exchange

Oleg Berzin
*CTO Office*
*Equinix*
Redwood City, CA, USA
oberzin@equinix.com

Rafael Ansay
*Edge Services*
*Equinix*
Singapore, Singapore
ransay@ap.equinix.com

James Kempf
*UC Santa Cruz Extension*
Santa Clara, CA, USA
kempf42@gmail.com[1]

Imam Sheikh
*Edelman Financial Engines*
Sunnyvale, CA, USA
imam.sheikh@gmail.com[1]

Doron Hendel
*Business Development*
*Equinix*
Redwood City, CA, USA
dhendel@equinix.com

*Abstract*—The IoT ecosystem suffers from a variety of problems around security, identity, access control, data flow and data storage that introduce friction into interactions between various parties. In many respects, the situation is similar to the early days of the Internet, where, prior to the establishment of Internet Exchanges, routing between different BGP autonomous systems was often point to point. We propose a similar solution, the IoT Exchange, where IoT device owners can register their devices and offer data for sale or can upload data into the IoT services of any of the big hyperscale cloud platforms for further processing. The goal of the IoT Exchange is to break down the silos within which device wireless connectivity types and cloud provider IoT systems constrain users to operate. In addition, if the device owner needs to maintain the data close to the edge to reduce access latency, the MillenniumDB service running in an edge data center with minimal latency to the edge device, provides a database with a variety of schema engines (SQL, noSQL, etc). The IoT exchange uses decentralized identifiers for identity management and verifiable credentials for authorizing software updates and to control access to the devices, to avoid dependence on certificate authorities and other centralized identity and authorization management systems. In addition, verifiable credentials provide a way whereby privacy preserving processing can be applied to traffic between a device and an end data or control customer, if some risk of privacy compromise exists.

## I. INTRODUCTION

While the Internet of Things (IoT) shows great promise for (among others) revolutionizing monitoring and control of industrial processes [1], enabling better urban management via "Smart Cities"[2], and delivering better and more affordable health care [3], a variety of problems stand in the way of realizing this promise. IoT requires combining devices having connectivity provided by different low power cellular protocols (such as LoRa [4], Cat-M1, and NB-IoT [5]) from different wireless providers, enterprise wireless networks based on WiFi, wired sensor networks, and the IoT platforms offered by hyperscale cloud providers, such as Microsoft's Azure IoT [6], where data from the devices is processed and stored and from which control protocol flows originate.

But simply integrating a collection of IoT devices from one owner running one wireless connectivity protocol registered to one network operator with one hyperscale cloud platform may not be enough to unlock the true value of the IoT ecosystem. For many reasons, enterprise customers often connect into multiple cloud providers and may want their data accessible across different cloud platforms. Device owners may have devices registered to multiple operators running different cellular IoT protocols, in addition to devices that use WiFi and wired connectivity. Some data may need to be processed with round trip latency constraints considerably tighter than with what can be achieved by a hyperscale cloud provider, in order to achieve proper control loop functioning.

Furthermore, each device network type has its own conventions about how to manage device registration. The hyperscale cloud IoT platforms use conventional centralized cloud identity and access management (IAM) systems for managing device identification and access control, where certificates or API keys prove authorization after the device has been registered into the cloud provider's IAM system. The result is a collection of siloed ecosystems with little interconnectivity between the silos and minimal ability to scale. As Romana, Zhoua and Lopez have observed:

> The existence of billions of heterogeneous objects also affects identity management. Beyond defining the actual scope of 'identity' in this context (e.g. underlying identity vs. real identity, core identity vs. temporary identity), we also need to provide some mechanisms for achieving universal authentication. Without authentication, it will not be possible to assure that the data flow produced by a certain entity contains what it is supposed to contain. Another important aspect related to authentication is authorization. If there is no access control whatsoever, everything will be accessed by everyone, which is neither viable nor realistic.[7]

The situation is not unlike the early days of the Internet, when fragmented network federations set up by different providers, largely for noncommercial use, increased the friction of connecting to and using the Internet [8]. Point to point connections between Internet Service Providers resulted in scalability problems that ultimately lead to the Internet Exchange Point (IXP). An IXP is a place where ISPs having a variety of business models can interconnect their networks through peering (cost-free exchange of traffic) [9], with the assumption that the volume of traffic exchanged between two peering partners is approximately equal over time.

---

[1]This work was completed while James Kempf and Imam Sheikh were at Equinix.

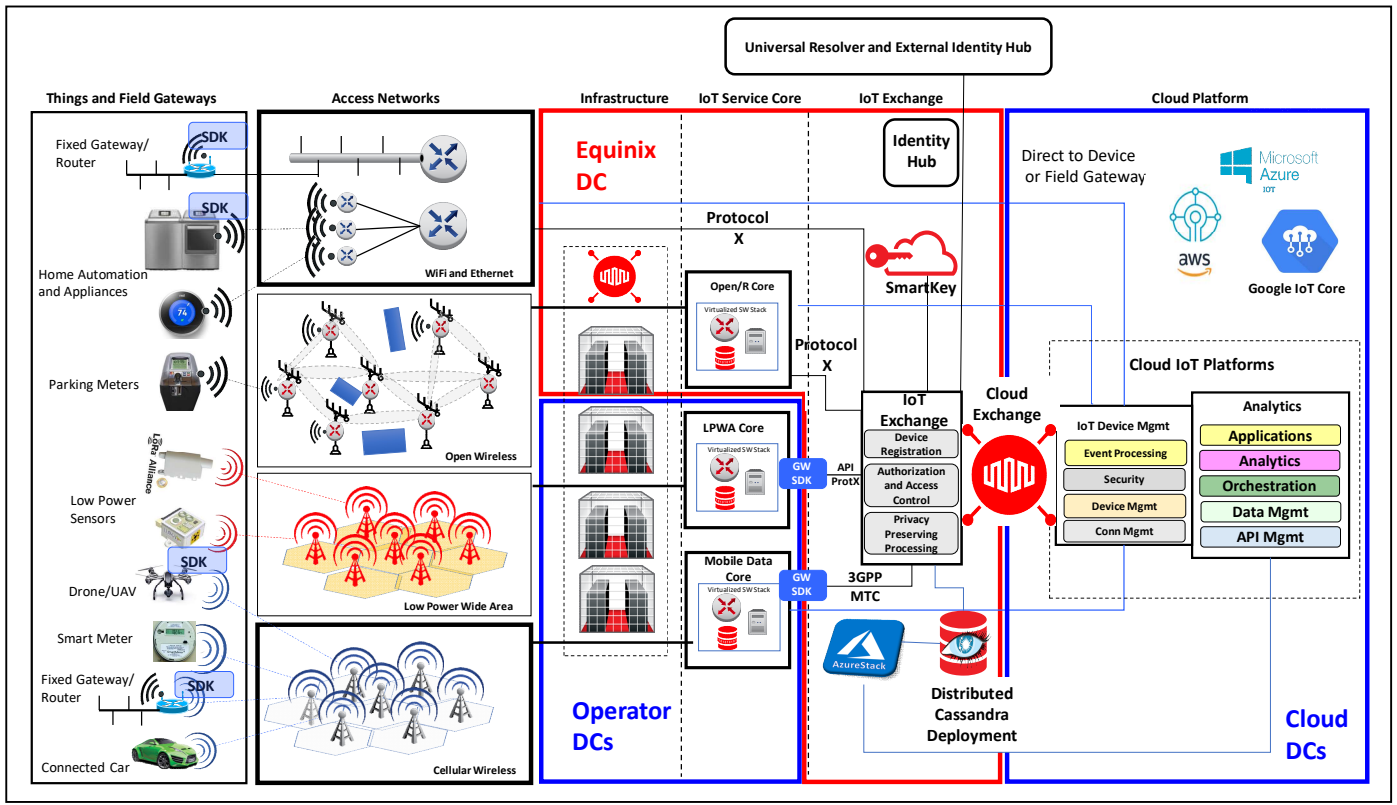

Fig. 1. The IoT Exchange Architecture in Context

We propose a similar solution for IoT, the IoT Exchange. The IoT Exchange provides a service where devices running different wired and wireless protocols and having different wireless network providers can register and exchange data with different hyperscale cloud provider IoT platforms, breaking down the silos between the device connectivity ecosystems and between the cloud providers' processing and storage services. Decentralized Identifiers [10], a new standard from the W3C Consortium for managing identity based on blockchain technology, provides identity management and authentication, while Verifiable Credentials [11], a developing standard from W3C for making cryptographically provable statements about a principal, provide authorization. We use DIDs and VCs for authorizing third party access to control and data to and from devices, for authorizing software update to the devices, and for controlling privacy when third party access to data may result in privacy compromise for people or organizations. We believe that the IoT Exchange can unlock the potential of IoT, just as IXPs did for the Internet, enabling the development of a market for IoT data.

The rest of this paper is organized as follows. In the next section, we present the overall architecture of the IoT Exchange. Section III briefly reviews DIDs and VCs, the underlying security technologies on which the IoT Exchange is built. In Section 0, we present how DIDs and VCs are used to achieve wireless network independent registration of devices, authorization of third parties to obtain data, privacy preservation, and authorized software update for IoT devices. Section V describes how we use a distributed Cassandra-compatible database to reduce latency edge-to-storage and provides a few screen shots of the IoT Exchange in operation. Finally, Section VI summarizes the paper and draws a few conclusions. To our knowledge, this work is the first to propose an integrated, cross wireless protocol, cross network operator, cross cloud platform for IoT.

## II. IOT EXCHANGE ARCHITECTURE

Fig. 1 illustrates the IoT Exchange Architecture in the context of the IoT ecosystem for controlling devices and collecting, storing and processing device data. To the left side of the IoT Exchange and moving right, the devices and their wireless access networks are shown. The architecture accommodates a broad variety of wired and wireless networks. Next to the right, the aggregation network is shown. This network consists of an Infrastructure layer and an IoT Service Core layer. An example of an IoT Service Core is the Evolved Packet Core (EPC) [12] for the 3GPP LTE 4G network, shown on in the bottom of the IoT Service Core box. Many Infrastructure and Service Core systems run in network operator data centers; in other cases, for example an industrial WiFi network or a LoRA network, the Infrastructure and Service Core may be deployed in an Equinix data center.

To the right of the IoT Exchange, the hyperscale cloud providers systems are shown. On a low level, these systems are responsible for event processing, connection security, connection management, and device management. On a high level, they provide the analysis and processing of non-latency sensitive data and API Management for external clients and Data Management for long term storage. They are connected into IoT Exchange through the Equinix Cloud Connect network. [13]

The IoT Exchange itself is shown in the middle and it provides three basic functions:
- Device registration across multiple wireless protocols and multiple operators,
- Authentication and authorized access control which includes authorized connectivity management,
- Privacy preserving pre-processing for control traffic and post-processing for data traffic to remove anything that might compromise privacy.

The IoT Exchange also provides connections to a low latency (toward the edge) database, which can be connected with cloud provider edge compute packages like Azure Stack [14] running on co-lo machines within the Equinix edge data center, for faster processing on low latency control loops.

Devices are identified in an operator and connectivity protocol independent way using DIDs. On the top, the IoT Exchange has connection to a local Identity Hub for maintaining DID documents locally if desired by the device owners, and to a Universal Resolver and any External Identity Hubs for obtaining DID documents of devices which have registered their DIDs remotely. DIDs are also used for identifying device owners, IoT connectivity customers and, indeed, the IoT Exchange itself are all identified by DIDs. We discuss the architecture of the DID resolution system in more detail in the next section. The IoT Exchange uses the Equinix SmartKey key management service [15] to store device private keys, removing the risk that an attacker may physically remove a device and thereby obtain access to the private key.

## III. DECENRALIZED IDENTIFIERS AND VERIFIABLE CREDENTIALS

Fig. 2 illustrates the DID architecture. DIDs [10] are Universal Resource Names (URNs) having the following format:

```
did:example:1234567890abcdefg
```

The `did` marks the URN as being of the decentralized identifier scheme while `example` indicates the DID method used to create the DID and thus the DID document type. The W3C DID standard supports extensibility by allowing new DID methods to be defined, and as of this writing, 41 methods are registered in the W3C registry [16]. In our prototype of the IoT Exchange, we are using the Microsoft ION method [17] which is based on the SideTree protocol [18]. Finally, the `1234567890abcdefg` is the actual DID itself, a unique identifier for the subject.

The DID document is a JSON formatted document containing the DID subject's public key and a specified verification method (a public key cryptosystem), which allows any receiving parties to verify that the DID does, in fact, belong to the party presenting the DID and also to authenticate any communication with that party. This definitively establishes the identity of the communicating party. Because the DID document is controlled by the party to which the DID refers via the private key, in contrast with public key certificates which are controlled by the Certificate Authority that issued them, DIDs are sometimes called "self-sovereign".

DIDs are resolved through the Universal Resolver [19]. The Universal Resolver acts like DNS but rather than resolving host names to IP addresses, it resolves DIDs to DID documents. Plug-ins for the various DID methods locate and return the DID documents for the DIDs that were created with the respective method. An Identity Hub is a globally accessible repository for DID documents and VCs. The IoT Exchange maintains an Identity Hub for devices registered with it, but device owners are free to choose an external Identity Hub if they so desire.

VCs are JSON formatted documents making cryptographically verifiable statements about a certain subject, usually identified by its DID. A VC contains the DID of the subject, in addition to other information related to the document like its creation date and how to prove that the document does indeed make a verifiable claim about the subject. The claims are included in the value of the `credentialSubject` property. For example, the following `credentialSubject` value is a JSON object asserting that the holder of the credential earned a BA in computer music from a French Canadian university identified by the `id` property:

```
"credentialSubject": {
    "id":
        "did:example:ebfeb1f712ebc6f1c276e12ec21",
    "degree": {
      "type": "BachelorDegree",
      "name": "<span lang='fr-CA'>
              Baccalauréat en musiques
              numériques</span>"
  }
}
```

For more about DIDs and VCs, please see the more complete review in [20].

## IV. IoT AUTHENTICATION AND AUTHORIZATION USING DIDs AND VCs

### A. Device Registration

The IoT Exchange uses DIDs to manage the identity of IoT devices. DIDs are also required for IoT device owners and IoT data/control customers but they need not register their

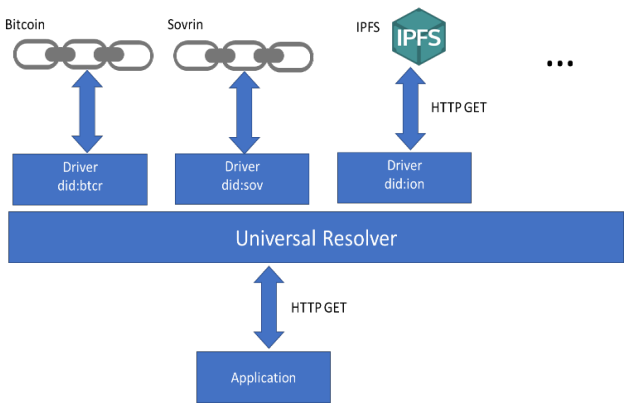

Fig. 2. DID Architecture (from [20])

DIDs through the IoT Exchange. The registration consists of two steps:
- A wireless/wired network specific enrollment step that registers the device on an operator network, or otherwise provides the device with a unique identifier tied to the specific connectivity context,
- The creation of a DID for the device that provides the device with a unique, global identifier.

Most operator-owned (cellular) wireless networks require some operator-specific and device-specific enrollment protocol to enroll a device into the operator network. For WiFi networks, or wired networks including those gatewayed from nonIP protocols such as BACnet [21] or CANbus [22], some identifier unique to the protocol, such as the MAC address for WiFi or Ethernet connected devices, can be used.

During DID registration, in addition to the standard properties, the DID document for the device records the connectivity-specific identifier as a `serviceEndpoint` for the connectivity-specific identifier. For example, here is what the `services` property looks like for an Ethernet connected device:

```
"services" : [{
   "id" : "did:example:1234567890abcdefg",
   "type":"EthernetMacAddress",
   "serviceEndpoint" : "00:0a:95:9d:68:16"
}]
```

This scheme for binding DIDs and their connectivity-specific identifiers mirrors the way IP addresses are bound to their link layer addresses in the Internet [23].

Once the device has been assigned a DID it can additionally use it for verified software update, as described in [20].

### B. Access Control and Privacy

#### 1) Parties and Privacy Access Control Lists

Once a device is registered to the IoT Exchange, access to the device is controlled by a VC. A customer wanting access to a collection of devices obtains the credential from the device owner, and then presents the IoT Exchange with the VC. The IoT Exchange verifies the credential and sets up access if the credential verifies. In the following discussion, agents operating on behalf of the principals play the following roles:

- The IoT Access Customer Agent takes a request from a customer or their agent (i.e. either a person or a machine) to access specific IoT devices periodically over a specified time period. For example, the request may be entered by a person through a browser-based Web interface provided by the IoT Exchange on behalf of the Device Owner.
- The IoT Device Owner Agent negotiates with the IoT Access Customer Agent for a VC to authorize device access.
- In some cases, the IoT Device Owner Agent may need to obtain public key signatures from other parties before issuing a VC authorizing access, for example the internal departments that manage the devices. The Authorizing Party Agents check a request against their policy, then sign a permission message to the IoT Device Owner Agent authorizing access.

In addition, the IoT Device Owner Agent and any Authorizing Parties maintain a list of DIDs, $L_p$, that are permitted to the devices without any privacy preserving pre/post processing. For example, a traffic light camera may compromise privacy by exposing the face of the driver and their license plate. Any traffic light camera data flows need to be sanitized of these identifying characteristics before being sent to customers that are not on $L_p$. An example of a DID that would not be on $L_p$ is the DID for the municipal police department, since they can view the full stream, as a law enforcement agency.

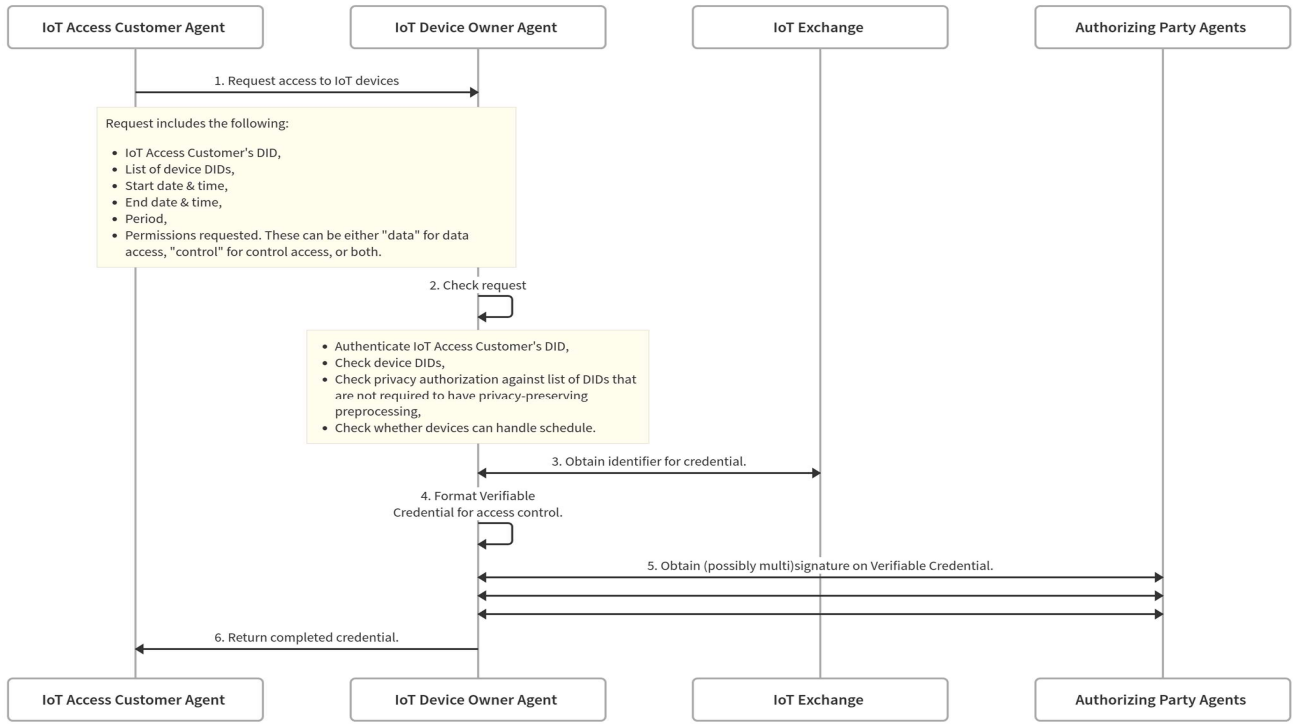

Fig. 3. Obtaining a VC to Access a Collection of Devices

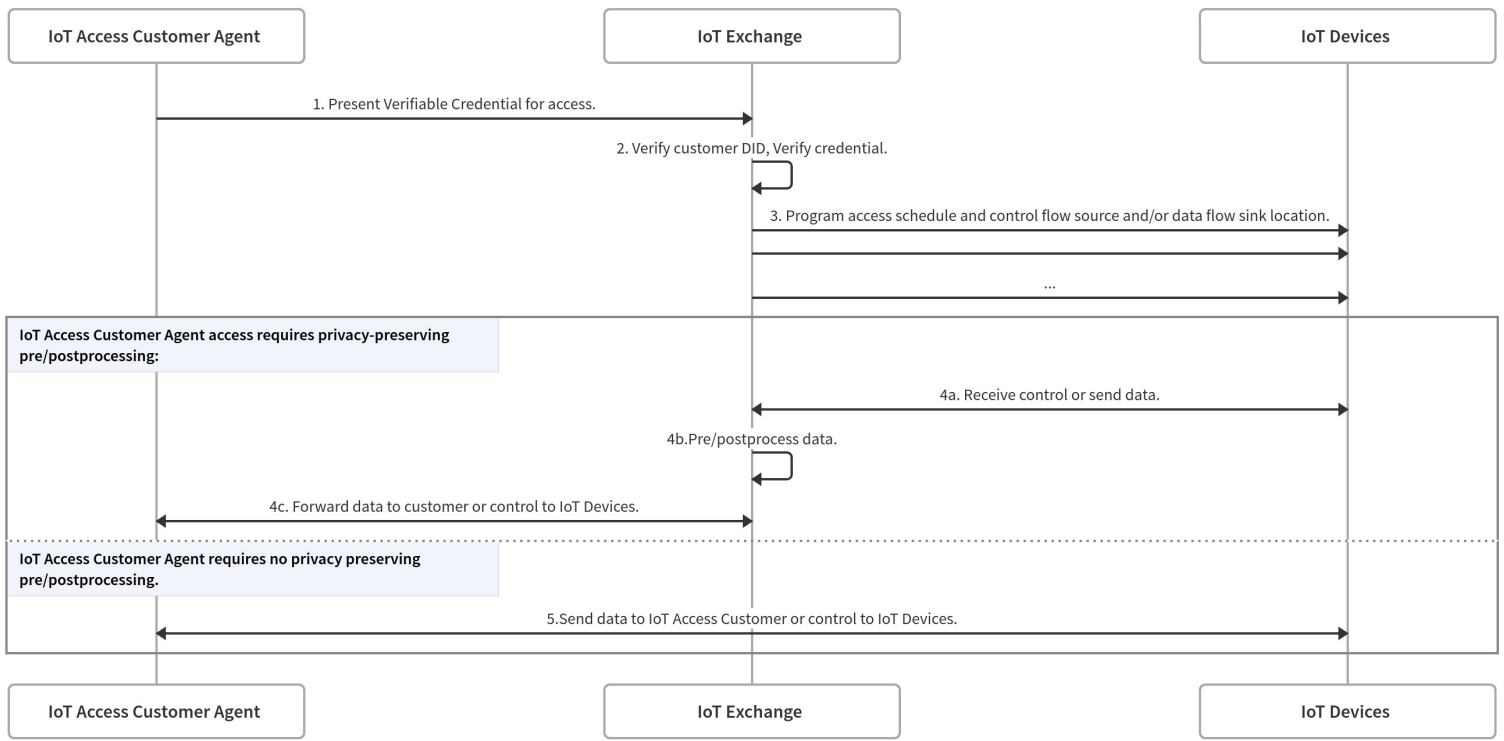

Fig. 4. Using a VC to Access A Collection of Devices

The IoT Device Owner Agent also maintains a pre/postprocessing specification, $S_p$, the entries of which contain a list of devices and a list of privacy preserving filters through which to run traffic. The details of what the filter removes are dependent on the device and what kind of data or control it generates or accepts. The flows from/to the device are run through each filter program in succession by the IoT Exchange prior to sending the data to the IoT Device Access Customer, for data access, or sending the control flow to a device on the DID list, for control access.

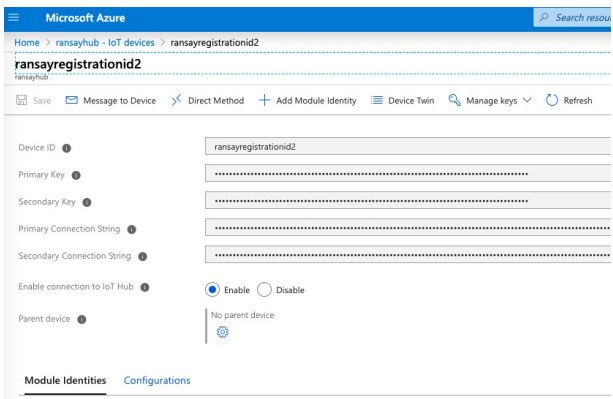

Fig. 5. Screen Shot of Azure IoT Registration with a DID

Finally, the Authorizing Parties maintain a policy list, $P_a$, of DIDs for organizations whose requests for either data accessor control access or both must be categorically denied by the IoT Device Owner Agent. This policy may be the result of specific rules put in place by the Authorizing Party or may be legal requirements from a larger legal jurisdiction, like a state's privacy law.

*2) Obtaining a VC to Access a Collection of Devices*

Fig. 3 shows the protocol flow for obtaining a VC to access a collection of devices. The IoT Device Customer Agent requests a VC from the IoT Device Owner Agent, providing its DID and a specification for the devices it wants to access, the beginning and ending period over which access is requested, the periodicity of the access, and the permissions, i.e. "control", "data" or both. The Owner Agent first verifies the Customer Agent's DID and the DIDs of the devices, then checks whether the devices can handle the load and whether the Customer Agent's DID is on the list of parties that can access the devices without privacy preserving processing.

The Owner Agent contacts the IoT Exchange to obtain an identifier for the VC. The identifier need not be a DID. Then the Owner Agent contacts any Authorizing Parties to determine if they have any policy constraints on the Customer Agent's access. If the Customer Agent DID is not on the list of DIDs exempt from privacy preserving processing, the `privacyPreserving` property in the VC is set to `true`. When everything is complete, the Owner Agent formulates the VC, signs it, and returns it to the Customer Agent. An example of the `credentialSubject` field in the VC is:

```
"credentialSubject": {
    "id":
        "did:example:ebfeb1f712ebc6f1c276e12ec21",
    "accessRequest" : {
        "deviceIds" : [ <list of device DIDs> ],
        "start" : "2019-10-01:00:00:00",
```

```
    "end" : "2019-10-30:23:59:59",
    "period" : "06:00:00",
    "permissions" : [ "data" ],
    "privacyPreserving" : true
  }
}
```

This allows the Customer Agent, whose DID is in the `id` field `data` only access to the list of devices in the `deviceIds` field from October 1, 2019 at midnight until October 31, 2019 one minute before midnight, at a period of every 6 hours, with privacy preserving postprocessing applied to the data stream if necessary.

*3) Using a Verifiable Credential to Request Device Access*
Figure 4 shows the protocol flow in which the Customer Agent uses the VC to request device access. The Customer Agent sends the VC to the IoT Exchange. The IoT Exchange verifies the credential, the Customer Agent's DID, the Owner Agent's DID, and the DIDs of the devices. It also checks if the VC identifier corresponds to an identifier that it has handed out to the Owner Agent. The IoT Exchange then programs the devices to allow the Customer Agent to access them.

Once access has been enabled, the IoT Exchange sets up a processing filter pipeline if the Customer Agent's `privacyPreserving` VC property is `true` and notifies the Customer Agent that the devices are ready for access. If the `privacyPreserving` VC property is `false`, the Customer Agent is notified without setting up the filter pipeline. Note that the messaging protocol, schema, and other details of the actual exchange between the device and the customer are either part of a public cloud provider's portal, listed on the Web page where the device access was requested, or need to be determined from some other source.

## V. THE IoT EXCHANGE IN OPERATION

Many IoT control loops are latency constrained, requiring communication, processing of incoming data, and formulation of a control response to be within a certain latency bound. Typical communication latency between an Equinix edge data center and an edge device runs around 20-40 ms, in contrast with latency to a centralized hyperscale data center which runs around 40-250 ms. A control loop with data located closer to the edge as a higher probability of hitting the latency bound. For this reason, the IoT Exchange includes a Cassandra compatible, distributed database that customers can use either with a cloud provider's on-prem stack, like Azure Stack, or with their own software running on a co-lo server. Control loops with latency bounds of less than 20 ms need to be deployed on far edge or on-prem hardware.

Fig. 5 shows a screen shot of the IoT Exchange integrated with Microsoft's IoT Azure IoT. IoT Exchange further extends device registration to the cloud by consuming the Azure IoT Hub Device Provisioning Service. In this way, the onboarded device has access to various Azure IoT Hub features such as monitoring and diagnostics logging as well as easy integration to other Azure cloud services such as the Azure Event Hub. IoT Exchange utilizes SmartKey key management service to maintain a mapping between the Azure keys, the device unique ID, the device DID, and the private-public key pair associated with the DID. This mapping is used to identify and verify data traffic flowing from the devices.

In Fig. 6, the results of privacy preserving v.s. no privacy preserving postprocessing on the data from a LoRa device that reports the device's location and the temperature. On the left, the device's location co-ordinates and LoRa id have been blurred out by the postprocessing filter. On the right, the id and location co-ordinates and id are visible.

## VI. SUMMARY AND CONCLUSIONS

In this paper, we have described and architecture and prototype implementation of the IoT Exchange, a system designed to enable interconnection between IoT devices using different connectivity protocols and network operators and connecting into different hyperscale cloud providers. The IoT Exchange utilizes decentralized identifiers and verifiable credentials to manage the identity of devices, device owners, and customers of device data and control, and verifiable credentials to control customer access to devices. In addition, devices can use DIDs to securely update their software. The private keys for DID creation are stored in the SmartKey key management service to avoid an attacker stealing the device and thereby gaining access to the key.

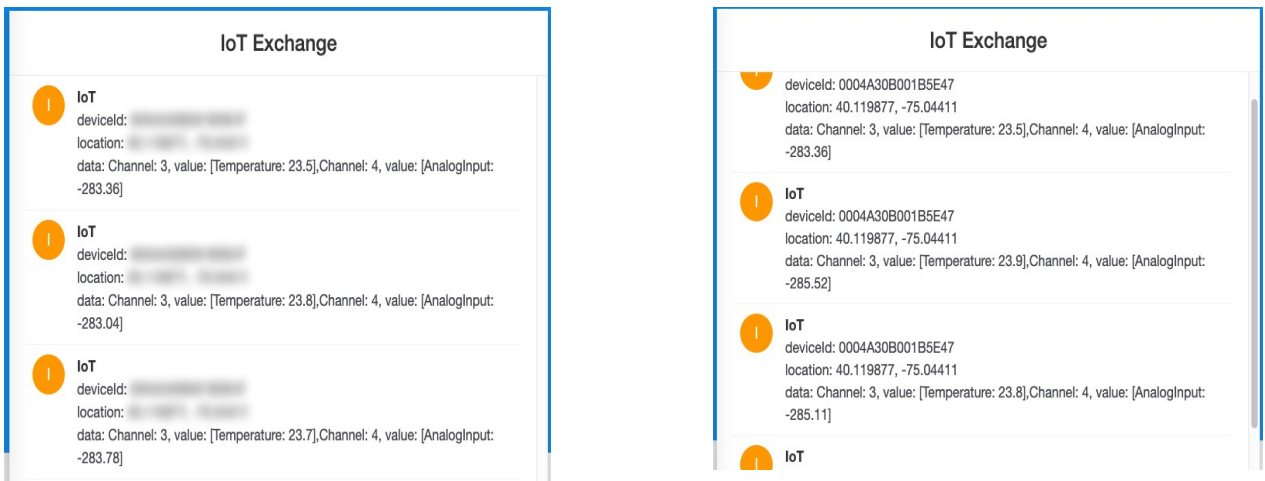

Fig. 6. Results of Privacy Preserving (left) v.s. no Privacy Preserving Postprocessing on a LoRa Device Data Stream

In addition to security support, the IoT Exchange includes the deployment of a distributed version of the Cassandra database within Equinix datacenters. The location of a data store within a 20-40 ms connectivity latency of edge devices enables tighter control loops than are possible thorough hyperscale cloud provider data centers. Connectivity into hyperscale cloud provider data centers is provided by Equinix Cloud Exchange Fabric for long term, bulk storage and other processing.

In summary, we believe that the IoT Exchange can unlock the potential value in IoT ecosystems in the same way that Internet Exchanges did for the early commercial Internet.